# Gap Junctions: The Claymore for Cancerous Cells

Masoud Asadi-Khiavi[1,2,3], Hossein Hamzeiy[1,2*], Sajjad Khani[2], Ailar Nakhlband[2] and Jaleh Barar[2,4*]

[1]Department of Pharmacology and Toxicology, School of Pharmacy, Tabriz University of Medical Sciences, Tabriz, Iran

[2]Research Center for Pharmaceutical Nanotechnology, School of Pharmacy, Tabriz University of Medical Sciences, Tabriz, Iran

[3]Student Research Committee, Tabriz University of Medical Sciences, Tabriz, Iran

[4]Ovarian Cancer Research Center, School of Medicine, University of Pennsylvania, Philadelphia, USA

**ABSTRACT**

***Introduction:*** Gap junctions play an important role in the cell proliferation in mammalian cells as well as carcinogenesis. However, there are controversial issues about their role in cancer pathogenesis. This study was designed to evaluate genotoxicity and cytotoxicity of Carbenoxolone (CBX) as a prototype of inter-cellular gap junction blocker in MCF7 and BT20 human breast cancer cells. ***Methods:*** The MCF7 and BT20 human breast cancer cell lines were cultivated, and treated at designated confluency with different doses of CBX. Cellular cytotoxicity was examined using standard colorimetric assay associated with cell viability tests. Gene expression evaluation was carried out using real time polymerase chain reaction (PCR). ***Results:*** MCF7 and BT20 cells were significantly affected by CBX in a dose dependent manner in cell viability assays. Despite varying expression of genes, down regulation of pro- and anti-apoptotic genes was observed in these cells. ***Conclusion:*** Based upon this investigation, it can be concluded that CBX could affect both low and high proliferative types of breast cancer cell lines and disproportionate down regulation of both pre- and anti-apoptotic genes may be related to interacting biomolecules, perhaps via gap junctions.

## Introduction

Cancer is a terrible disorder facing human today and unfortunately, lack of deep knowledge about cancer biology causes perfect treatment failing. However, several strategies have been developed upon recent findings in molecular biology. One of these approaches appears to be valuable because of inhibiting inter-cellular gap junction communication (GJIC) between cells. As a rule, lack of gap junctions was commonly observed in carcinogenesis (Leithe *et al.* 2006). Some critical small biomolecules like cAMP and $Ca^{2+}$ as well as several small sized amino acids could pass through this communicative channel. Cellular apoptosis pathway, Akt signaling pathway, EGF pathway, p53 pathway, inhibition of matrix metalloproteinases (MMPs) and VEGF family ligands and receptors are the most important biomolecules and ways in cancer biology which could be affected by these small agents. As mentioned, EGF pathway and its obedience biomolecules has vital role in carcinogenesis of epithelial derived carcinomas like breast and lung cancers (Cameron *et al.* 2003). This cell-surface tyrosine kinase receptor is a member of the ErbB family of receptors and mutations which could result in some epithelial derived carcinomas including breast cancer through affecting its expression or activity. EGFR over expression was reported in 20% - 50% of breast cancer (Agelaki *et al.* 2009). EGFR is activated by binding its specific ligands, including epidermal growth factor (EGF) and transforming growth factor α (TGFα). Due to activation by its ligands, EGFR switches from an inactive monomeric form to an active homodimer (Dong J and Wiley HS 2000). Several signal transduction cascades was initiated by downstream signaling proteins, principally the mitogen-activated protein kinase (MAPK), phosphatidylinositol-3-kinase (PI-3K/Akt), and finally signal transducer and activator of

*Corresponding author:* Hossein Hamzeiy (PhD) and Jaleh Barar (PhD), Tel.: + 98 411 3367914, Fax: +98 411 3367929,
E-mail: hhamzeiy@hotmail.com, jbarar@tbzmed.ac.ir

transcription (STAT) pathways (Loew *et al.* 2009; Afaq *et al.* 2008; Camp *et al.* 2005; Mamot and Rochlitz. 2006) which mainly undergoes to cell proliferation, adhesion and migration (Agelaki *et al.* 2009) as well as involvement in the metastasis, angiogenesis and apoptosis inhibition (Gialeli *et al.* 2009; Camp *et al.* 2005). In regards to these miscellaneous biomolecules, it is believed that EGFR signal transduction could be modulated by potent small molecules which may be transferred via gap junctions and could prevent or promote pathogenesis of cancer (Nicholson 2003). CBX ($C_{34}H_{50}O_7$) [3-(3-carboxy-1-oxopropoxy)-11-oxo-Olean-12 en-29-oic acid] (Fig. 1) as prototypic inter-cellular communication blocker which mediated via Gap junctions, inhibits cell growth and induces apoptosis in several types of cancer. CBX is a semi-synthetic derivation of Glycyrrhizin obtained from Licorice root and demonstrates a superficial resemblance to steroid structures. This drug is often used as a gap junction inhibitor in cells (Ye, Z. C., *et al.* 2003; Li, J., *et al.* 2001). However, there are some controversial issues about gap junction role in carcinogenesis(Cronier, Crespin *et al.* 2009). This study was designed to determine the impacts of CBX on MCF7and BT20 human breast cancer cells in the view of cell growth and apoptosis particularly at mRNA transcriptomics level.

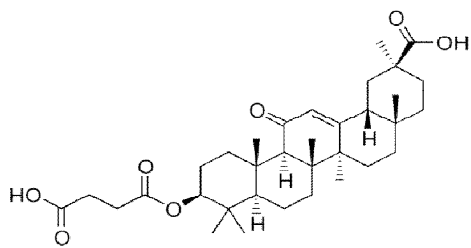

**Fig. 1.** The structure of carbenoxolone (CBX).

## Materials and methods

### Materials
The MCF7 and BT20 human breast cancer cell lines, Cell culture plates and flasks were obtained from National cell bank of Iran (Pasteur institute, Iran) and IWAKI, Japan respectively. RPMI1640 medium, Trypsin-EDTA (0.05%–0.02%), Tri-Reagent, Dithiotheritol (DTT), first strand reverse transcriptase (RT) buffer, Moloney murine leukemia virus reverse-transcriptase (MMLV-rt, 200U/μl) was purchased from Sigma Aldrich Co., (Poole, UK). Agarose and Fetal bovine serum (FBS) were bought from Gibco, Invitrogen (Paisley, UK). Ribonuclease inhibitor was obtained from Fermentas, Canada. *STAT1, AKT1, GAPDH (Glyceraldehyde 3-phosphate dehydrogenase), TP53,* *BCL2 (B-cell lymphoma 2)* and *CYCS (Cytochrome c)* Forward and reverse oligonucleotide primers were provided from MWG-Europhin, (Ebersberg, Germany). The SYBR Green PCR Master Mix was obtained from Applied Biosystems, Foster City, USA. Diethyl pyrocarbonate (DEPC) treated water; streptomycin and Penicillin G were obtained from CinnaGen, (Tehran, Iran). Taq DNA polymerase, polymerase chain reaction (PCR) reaction buffer, Random hexamer primers (pdN6), deoxynucleotide triphosphate monomers (dNTPs) and MgCl2 were obtained from QIAGEN (Crawley, UK). Carbenoxolone was obtained from Sigma Aldrich Co., (Poole, UK). All other chemicals not mentioned above were in the highest quality available.

### Cell culture
The MCF7and BT20 human breast cancer cells were seeded on 6-well plates at a density of $4.0 \times 10^4$ cells/cm2. These cells were saved in a humidified incubator (95% air and 5% CO2) at 37°C. Cell culture media consisted of RPMI 1640 complemented with 10% FBS, penicillin G (200 U/ml) and streptomycin (200 μg/ml). CBX was solved in RPMI 1640 media at concentrations of 50μM, 100μM, 150μM, 300μM, 600μM and finally 1000μM (1Mm) and then cells were lay open to viability examinations as well as gene expression profiling using real time RT-PCR.

### Cell viability assessment
The cultured MCF7and BT20 human breast cancer cells at 40–50% confluency were subjected to MTT assay in the 96-well plates. The cells were treated with a range of CBX concentrations and each group were incubated for 24 and 48 hours at 37°C. After which, the cells of each group were washed once with phosphate buffered saline (PBS) and culture medium was replaced with 150 μl fresh media and then 50 μl MTT reagent (2 mg/ml in PBS) was added to each well. After a 4 hr incubation with MTT at 37°C, medium was removed and the cells were exposed to 200 μl DMSO and 25 μl of Sorenson buffer (0.1 M glycine, 0.1 M NaCl, pH 10.5). The cultures were incubated for 30 min at 37°C for formazan crystals dissolving and then UV absorbance was measured at 570 nm using a spectrophotometric plate reader, ELx 800 (Biotek, CA, USA). Trypan blue exclusion test was performed for growth inhibition assessment of CBX. Briefly, cells were counted based on the ability of cells to exclude trypan blue dye using a hemocytometer for obtaining vital cell numbers versus death cell numbers. Light microscopic examinations were conducted for morphological assessments using Olympus invert microscope CKX41 equipped with Olympus DP20 camera and CellA 3.3 software (Olympus optical Co., Ltd., Tokyo, Japan).

*Gene expression analysis*
Cells were subjected to gene expression evaluation just at 150μM which was obtained as IC50 of CBX at the exposure time of 24 hr and total RNA was isolated using TriReagent according to manufacturer guideline. The quality and quantity of the extracted RNA was measured respectively by electrophoresis and NanoDrop®1000 Spectrophotometer (Nano- Drop Technologies, Wilmington, DE, USA). Primers were designed from published Gene Bank sequences using Beacon Designer 5.01 (Premier Biosoft International, http://www.premierbiosoft.com) and listed in Table 1. The reverse transcription reaction was performed using 11 μl DEPC treated water, 4 μl MMLV buffered with DTT, 2 μl dNTPs (10 mM), 0.5 μl pdN6 (400 ng/μl), 1 μg of RNA, 0.5 μl RNasin (40 U), and 1 μl MMLV-rt (200 U) in a total volume of 20 μl. Thermal cycler program was: 95 °C for 5 min before addition of RNasin and reverse transcriptase, and incubation at 25 °C for 10 min, 42 °C for 50 min and then they were subjected to real time PCR examinations. PCR experiment was performed to assess the expression of selected key genes as mentioned above.

All amplification reactions were performed in a total 25μl volume using the iQ5 Optical System (Bio-Rad Laboratories Inc., Hercules, USA). Each well contained: 1 μl cDNA, 1 μl primer (100 nM each primer), 12.5 μl 2X Power SYBR Green PCR Master Mix and 10.5 μl RNAse/DNAse free water. Thermal cycling conditions were as follows: 1 cycle at 94°C for 10 min, 40 cycles at 95°C for 15 sec, 56-62°C for 30 sec, and 72°C for 25 sec. Interpretation of the result was performed using the Pfaffle method and the CT values were normalized to the expression rate of *GAPDH* as a housekeeping gene (Pfaffl 2001). All reactions were performed in triplicate and negative control included in each experiment as well as internal positive control.

*Data analyzing*
One way ANOVA was performed for statistical analysis using SPSS version 18 following with Tukey's Post Hoc Test. A *p value* less than 0.05 was considered to statistical differences representation. MTT assays and Trypan blue exclusion tests were performed in at least five independent experiments and gene expression evaluation using real time RT-PCR performed in triplicate manner.

**Results**

*Cytotoxicity assessment*
Trypan blue exclusion test and MTT assay were conducted to evaluate the cytotoxicity of carbenoxolone in MCF7 and BT20 human breast cancer cells. As shown in Fig. 2, CBX treated cells led to growth inhibition in dose dependent manner ($p < 0.05$).

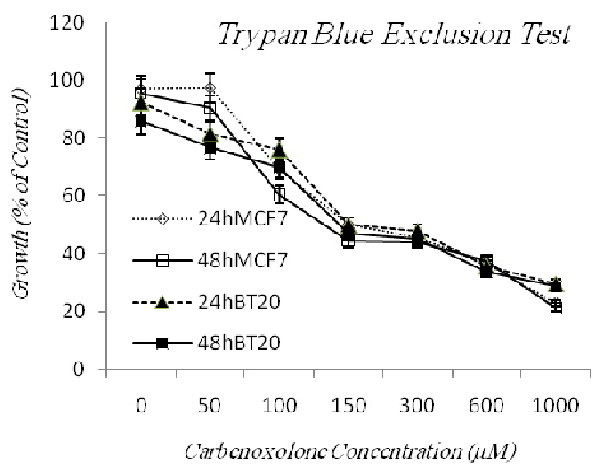

**Fig. 2.** The Effects of CBX on growth inhibition of MCF7 and BT20 cells. Cells were exposed to the indicated concentration (50-1000μM) for 24 h (♦) and 48 h (□) in the case of MCF7 as well as for 24 h (▲) and 48 h (■) in the case of BT20. The number of viable cells was determined by Trypan blue exclusion test. Growth inhibition in each treatment was expressed as a percentage of the control. Each value represents the mean ± SE of at least five independent experiments.

**Table 1.** Genes with corresponding primers which used for gene expression analysis

| Gene name | Accession No. | Sequence | Product length |
|---|---|---|---|
| GAPDH | NM_002046.3 | F: 5'-AAGCTCATTTCCTGGTATGACAACG-3'<br>R: 5'-TCTTCCTCTTGTGCTCTTGCTGG-3' | 126 |
| BCL2 | NM_000633.2 | F: 5'-CATCAGGAAGGCTAGAGTTACC-3'<br>R: 5'-CAGACATTCGGAGACCACAC-3' | 181 |
| AKT1 | NM_005163.2 | F: 5'- CGCAGTGCCAGCTGATGAAG -3'<br>R: 5'- GTCCATCTCCTCCTCCTCCTG -3' | 185 |
| CYCS | NM_018947.4 | F: 5'- ACCTTCCATCTTGGCTAGTTGTG-3'<br>R: 5'- ATCGCTTGAGCCTGGGAAATAG-3' | 129 |
| STAT5 | NM_007315.3 | F: 5'- TCATCAGCAAGGAGCGAGAG-3'<br>R: 5'- TCAGGGAAAGTAACAGCAGAAAG-3' | 196 |
| TP53 | NM_001126114 | F: 5'-TCAACAAGATGTTTTGCCAACTG-3'<br>R: 5'-ATGTGCTGTGACTGCTTGTAGATG-3' | 118 |

F: Forward; R: Reverse

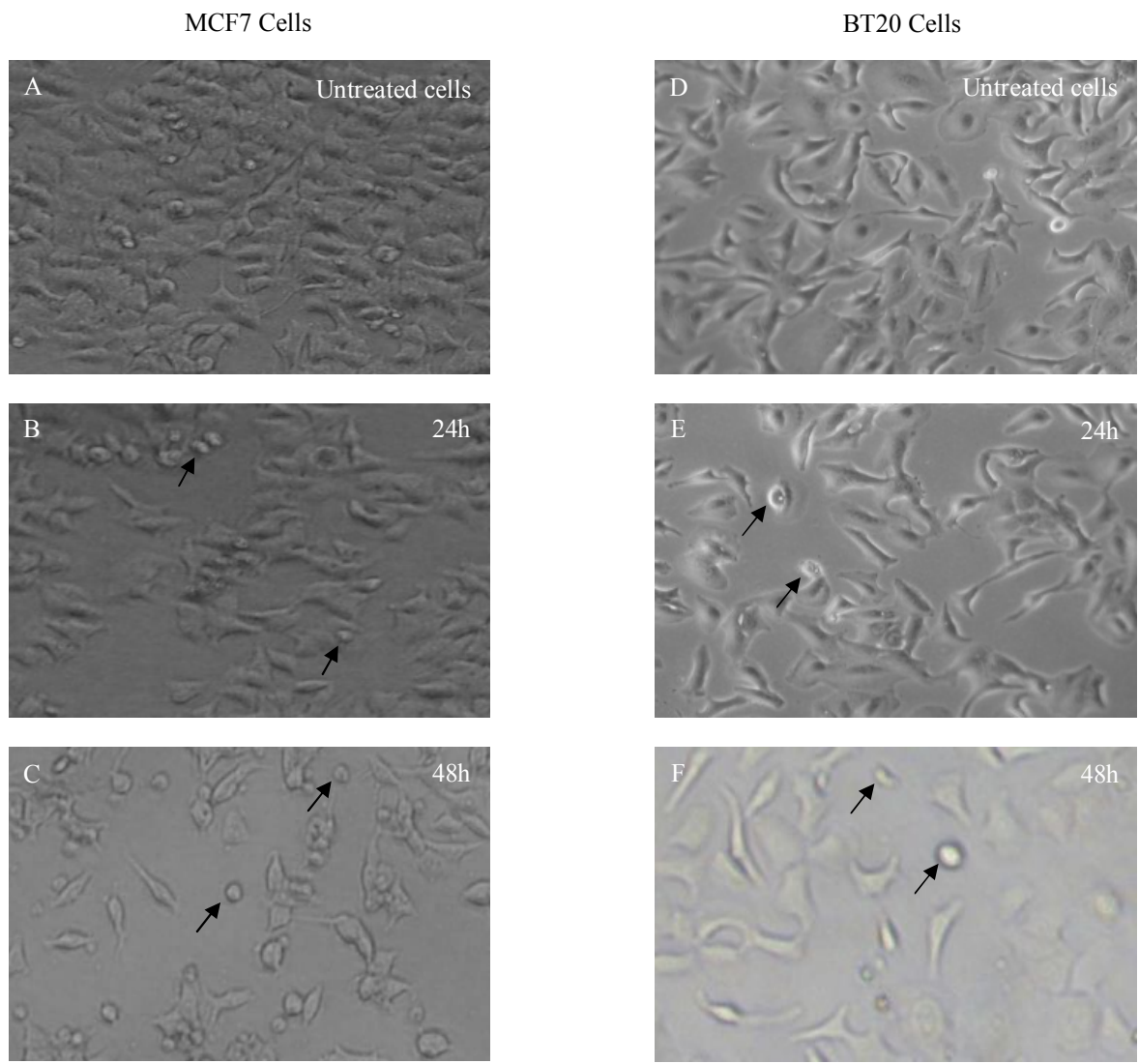

**Fig. 3.** Photomicrographs of MCF7 and BT20 cells were taken by a light microscope at a magnification of 40×. MCF7 and BT20 cells were treated by 150 μM of CBX. The death cells are shown by black arrows. Panel A represent untreated MCF7 cells versus CBX treated MCF7 cells in panels B and C at exposure times of 24h and 48h respectively. Panel D represent untreated BT20 cells versus CBX treated BT20 cells in panels E and F at exposure times of 24h and 48h respectively.

As it shows, the growth of treated BT20 cells was affected notably, but in the case of MCF7 cells, anti-proliferative effect of CBX was lower than previous cells. However, CBX also showed statistically significant cytotoxicity in the MCF cells. Additionally, the treated and untreated cells displayed distinct morphologic differences in normal and death cells number and appearance upon light microscopic examinations (Fig. 3). Regarding the obtained data from MTT experiment as a standard colorimetric cell viability assay, IC50 were calculated as 150 μM in both cells (Fig 4), which is relatively similar to obtained IC50 from data of growth inhibition evaluation via Trypan blue exclusion test in Fig. 2. In the view of statistics (mean ±SE of at least five independent experiments), reduction in number of cells was observed after treatment by 150 μM of CBX as same as reduction in total amount of extracted RNA in both cells (data not shown). Despite mild exacerbation in CBX mediated cytotoxicity in a time dependent manner, it was not statistically significant even at exposure time of 72 hr in both cell lines (data not shown).

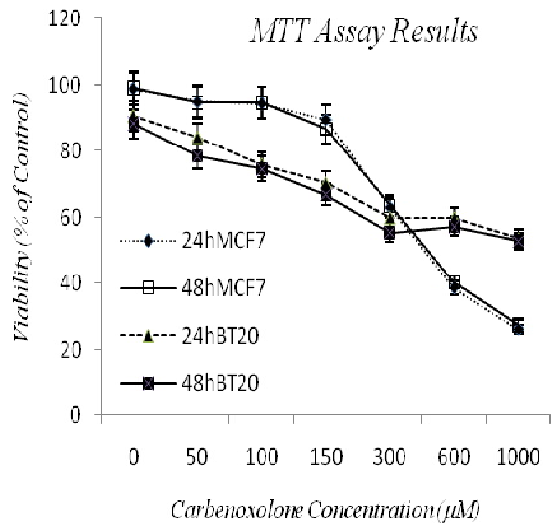

**Fig. 4.** Results of MTT assay for evaluation of cell viability. Cells were exposed to the indicated concentration of CBX (50-1000µM) at exposure time of 24 h (♦) and 48 h (□) and also are repeated in the case of BT20 for mentioned concentrations and exposure times. As shown in this diagram, CBX in 150 µM of concentration could significantly suppress the production of formazan as reductive metabolite of MTT (3-(4, 5-Dimethylthiazol-2-yl)-2, 5-diphenyltetrazolium bromide) in both cell lines. Each value represents the mean ± SE of at least five independent experiments.

*Gene expression analyses*

MCF7 and BT20 cells were incubated with CBX (150 µM) for 24 h. Then expression levels of candidate genes which mentioned above were detected by using real time PCR and Interpretation of the results was performed using the Pfaffle method and finally the CT values of each gene were normalized to the expression behavior of *GAPDH* as a housekeeping gene. All reactions were performed in triplicate and negative control included in each experiment and then quantified as a view of SYBR Green emitted fluorescent intensity.

We performed a quantitative PCR approach to assess the effect of carbenoxolone on the expression of apoptosis related genes *BCL2* and *CYSC*, EGF pathway candidate genes *AKT1* and *STAT5* and finally *TP53* as crucial gene the cell cycle control and a tumor suppressor gene which is involved in cancer preventing (May, P. and May, E. 1999). The mRNA expression of these genes versus *GAPDH* expression is illustrated in Figure 5. The respective ratio of expressed gene over *GAPDH* expression was used to represent gene expression changes and was demonstrated quantitatively in Fig. 5. Normalization of our data to the expression rate of *GAPDH* as a housekeeping gene, revealed significant down regulation of all mentioned genes in both type of cells but there was not significant changes among these candidate genes themselves.

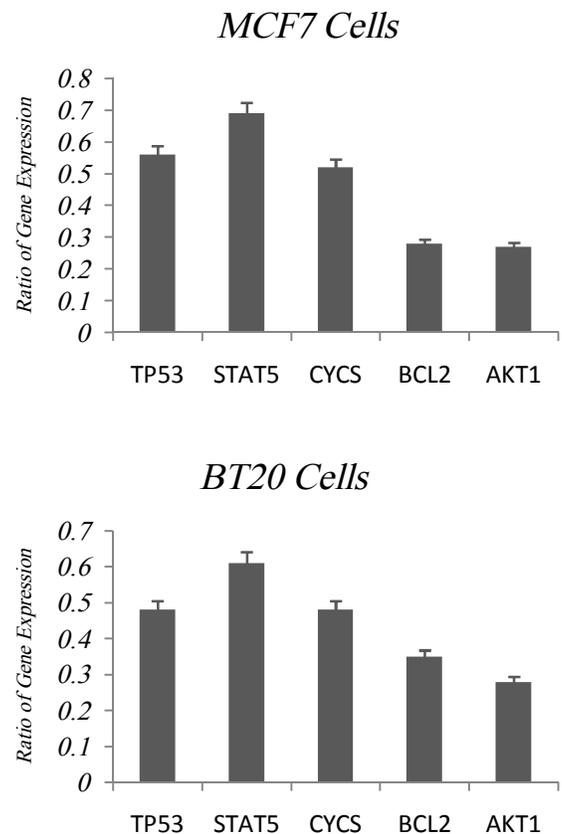

**Fig. 5.** Quantitative real time PCR assessments of CBX on MCF7 cells (Panel A) and BT20 cells (Panel B). Normalized relative expressions intensity ratio of the genes upon the expression of housekeeping *GAPDH* gene. All of the genes expressions are significant statistically upon *GAPDH*. The values were obtained as a view of SYBR Green emitted fluorescent intensity in triplicate experiments statistically (mean ± SEM).

**Discussion**

Although, various drugs have been proposed for the treatment of breast cancer, lack of deep knowledge about cancer biology and adverse side effects of routine anticancer drugs are obstacles of complete treatment. Consequently, many efforts are in progress to find new approaches to overcome cancer. Inhibition of gap junctions is one of these efforts and CBX is a prototypic inhibitor of them. However, in the case of CBX, total amount of extracted RNA decreased in treated group in both types of cells (data not shown) which could indicate the reduction of cell numbers after incubation with drug. It should be highlighted that MCF7 cells have more proliferative and progressive behavior, dependent on their EGF, progesterone and estrogen receptors. MTT assay revealed significant reduction in cell viability in BT20 cells (Fig. 4), while MCF7 did not show such effects due to mentioned supportive receptors in MCF7 cells (Rulli, Antognelli *et al.* 2006). According to the results, CBX inhibits MCF7 and BT20 cells growth and viability just in a dose dependent manner with IC$_{50}$ values of 150 μM after 24 h of exposure. The anti-ploriferative effects of CBX have been reported in other cancer cells (Kawashima *et al*. 2009). Apoptotic effects of CBX have been also reported in some types of cells (Pivato *et al*. 2006; Waddell *et al*. 2000). Among several mechanisms of action which have been proposed for pharmacological activity of CBX, induction of Mitochondrial Permeability Transition (MPT) and release of cytochrome c is prominent (Salvi *et al*. 2005; Pivato *et al*. 2006). Therefore, it can be considered as an apoptosis stimulating drug (Isbrucker *et al*. 2006; Salvi *et al*. 2005). Due to structural similarity to glucocorticoids, CBX inhibits Hydroxysteroid dehydrogenases (HSD) in cancer cells and induces apoptosis via interaction with glucocorticoid and mineralocorticoid receptors (Greenstein *et al*. 2002; Voutsas *et al*. 2007). It seems that *TP53* and *CYCS* genes as well as *STAT1* gene down regulation are compensatory responses against anti-apoptotic genes down regulation which totally arise from stress oxidative effect of Carbenoxolone (Farczadi, Kaszas *et al*. 2002). *AKT1* and *BCL2* genes are anti-apoptotic down regulated genes with similar effect on cancer cells. Of these down regulated genes, the *BCL2* gene encodes the bcl2 protein that involved in a wide variety of cellular activities such as homeostasis and tumorigenesis. The encoded protein is able to reduce the release of pro-apoptotic cytochrome c from mitochondria and block caspase activation. This suggests a cytoprotective and cell survival functionality for this gene as reported previously ( Karsan *et al*. 1996; Choi *et al*. 1995)  but as it mentioned, cytochrome c has been down regulated in CBX group, probably due to carbenoxolone stress oxidative effect (Salvi *et al*. 2005; Pivato *et al*. 2006) and *BCL2* counter regulatory gene, was not able to be compensated. In the view of transcryptomics, it seems that Carbenoxolone affects cancer cell viability despite some ambiguous findings in cell viability and cytotoxicity assays. Lewenstein and his co-workers provided the "Homologous Growth Control (HGC)" theory approximately forty  years ago which talks about gap junctions' delirious role in carcinogenesis particularly in the case of gap junction intercellular communication(GJIC) between normal and cancer  cells (Cronier, Crespin *et al*. 2009) and CBX has confirmed inhibitory effects on gap junction-mediated intercellular communication (GJIC) particularly in attached cells as well (Winmill *et al*. 2003; Paraguassu-Braga *et al*. 2003). Therefore we cannot establish a precise mechanism for CBX activity.

## Conclusion

Breast cancer has emerged as a leading cause of cancer death in the world particularly in women. Unfortunately, the current therapeutic strategies are ineffective in some cases; hence, there exists an increasing demand for more effective therapies. We examined the CBX in this matter. In conclusion, based on the results obtained in this study, we suggest that CBX be favourable for *in vivo* applications; nevertheless further investigations are yet to be fulfilled to authenticate its *in vivo* usefulness. This obviously means that there is a novel relationship between cell-cell communications and inter-cellular cross talking, perhaps via gap junctions, at whom we propose that interestingly orchestrated biochemical machinery of cells may be related to some downstream cell signalling which may play a key role in development of cancer. On the other hand, the main question is remained: Are gap junctions friend or enemy? Therefore we propose evaluation of CBX effects on genetically modified cancer cell lines particularly in transcriptomics and functional genomics level using DNA chip technology.

**Ethical issues**
None to be declared.

**Conflict of interests**
The authors declare no conflict of interests.

**Acknowledgement**
Authors would like to thank Research Center for Pharmaceutical Nanotechnology (RCPN), Tabriz University of Medical Sciences, Tabriz, Iran for the financial support of this project (grant No. 87011) that is a part of PhD thesis number 32 Also authors are thankful to Dr Y. Omidi for his kind support.